\documentclass[sigconf]{acmart}

\usepackage{multirow}
\usepackage{tabularray}
\usepackage{tabularx}
\usepackage[normalem]{ulem}
\useunder{\uline}{\ul}{}
\usepackage{makecell}
\usepackage{graphicx}
\usepackage{subfigure}
\newtheorem{myDef}{Definition}
\usepackage[ruled]{algorithm2e}
\usepackage{mathalfa}
\usepackage{amsmath}
\usepackage{caption}
\usepackage{array}
\usepackage{enumitem}
\usepackage{verbatim}
\usepackage{flushend}

\AtBeginDocument{%
  \providecommand\BibTeX{{%
    \normalfont B\kern-0.5em{\scshape i\kern-0.25em b}\kern-0.8em\TeX}}}

\begin{document}

\title{Consistency and Discrepancy-Based Contrastive Tripartite Graph Learning for Recommendations}


\author{Linxin Guo}
\email{guolinxin@cqu.edu.cn}
\affiliation{%
  \institution{Chongqing University}
  \city{Chongqing}
  \country{China}
}

\author{Yaochen Zhu}
\email{uqp4qh@virginia.edu}
\affiliation{%
  \institution{University of Virginia}
  \city{Charlottesville}
  \country{USA}
  }

\author{Min Gao}
\authornote{Corresponding author}
\email{gaomin@cqu.edu.cn}
\affiliation{%
  \institution{Chongqing University}
  \city{Chongqing}
  \country{China}
}

\author{Yinghui Tao}
\email{taoyinghui@cqu.edu.cn}
\affiliation{
  \institution{Institute of Guizhou Aerospace Measuring and Testing Technology}
    \city{Guiyang}
    \country{China}
  }

\author{Junliang Yu}
\email{jl.yu@uq.edu.au}
\affiliation{%
  \institution{the University of Queensland}
  \city{Queensland}
  \country{Australia}
  }

\author{Chen Chen}
\email{chenannie45@gmail.com}
\affiliation{%
  \institution{University of Virginia}
  \city{Charlottesville}
  \country{USA}
}

\renewcommand{\shortauthors}{Linxin Guo et al.}


\begin{abstract}
Tripartite graph-based recommender systems markedly diverge from traditional models by recommending unique combinations such as user groups and item bundles.   
Despite their effectiveness, these systems exacerbate the long-standing cold-start problem in traditional recommender systems, because any number of user groups or item bundles can be formed among users or items. To address this issue, we introduce a \underline{C}onsistency and \underline{D}iscrepancy-based graph contrastive learning method for tripartite graph-based \underline{R}ecommendation (CDR). 
This approach leverages two novel meta-path-based metrics—consistency and discrepancy—to capture nuanced, implicit associations between the recommended objects and the recommendees. 
These metrics, indicative of high-order similarities, can be efficiently calculated with infinite graph convolutional networks (GCN) layers under a multi-objective optimization framework, using the limit theory of GCN. 
Additionally, we introduce a novel Contrastive Divergence (CD) loss, which can seamlessly integrate the consistency and discrepancy metrics into the contrastive objective as the positive and contrastive supervision signals to learn node representations, enhancing the pairwise ranking of recommended objects and proving particularly valuable in severe cold-start scenarios. Extensive experiments demonstrate the effectiveness of the proposed CDR. The code is released at \url{https://github.com/foodfaust/CDR}.
\end{abstract}

\ccsdesc[500]{Information systems~Recommender systems}

\keywords{Tripartite Graph-Based Recommendations, Extreme Cold-Start, Consistency and Discrepancy}

\copyrightyear{2024}
\acmYear{2024}
\setcopyright{acmlicensed}
\acmConference[KDD '24] {Proceedings of the 30th ACM SIGKDD Conference on Knowledge Discovery and Data Mining }{August 25--29, 2024}{Barcelona, Spain.}
\acmBooktitle{Proceedings of the 30th ACM SIGKDD Conference on Knowledge Discovery and Data Mining (KDD '24), August 25--29, 2024, Barcelona, Spain}
\acmISBN{979-8-4007-0490-1/24/08}
\acmDOI{10.1145/3637528.3672056}

\maketitle

\section{Introduction}
With the rapid development of the Internet, recommender systems have become crucial in online platforms \cite{lu2012recommender,cheng2016wide,10.1145/3447395}. Traditional systems, modeling the interactions between user and item, have achieved significant results in both industry and academia \cite{rendle2012bpr,he2020lightgcn,wang2019neural,zhu2022mutually}. 
Recently, recommender system research has embraced a trend of formulating users, items, and their interactions as a bipartite graph, prompting the development of bipartite graph-based methods like NGCF \cite{chen2019collaborative} and LightGCN \cite{he2020lightgcn}, which have notably advanced the field. 
However, tripartite graph-based recommendation, involving interactions among users, items, and user groups/item bundles (e.g., group recommendations \cite{guo2020group,sankar2020groupim,zhang2021double} and bundle recommendations \cite{9546546,chen2019matching,10.1145/3534678.3539229,10.1145/3022185}), has recently emerged and garnered substantial research interest due to its unique interaction model among three heterogeneous entities/nodes \cite{cao2018attentive, ghaemmaghami2021deepgroup, pathak2017generating, bai2019personalized}. 

\begin{figure}[ht]
    \centering
    \includegraphics[width=1\linewidth,height=0.14\textheight]{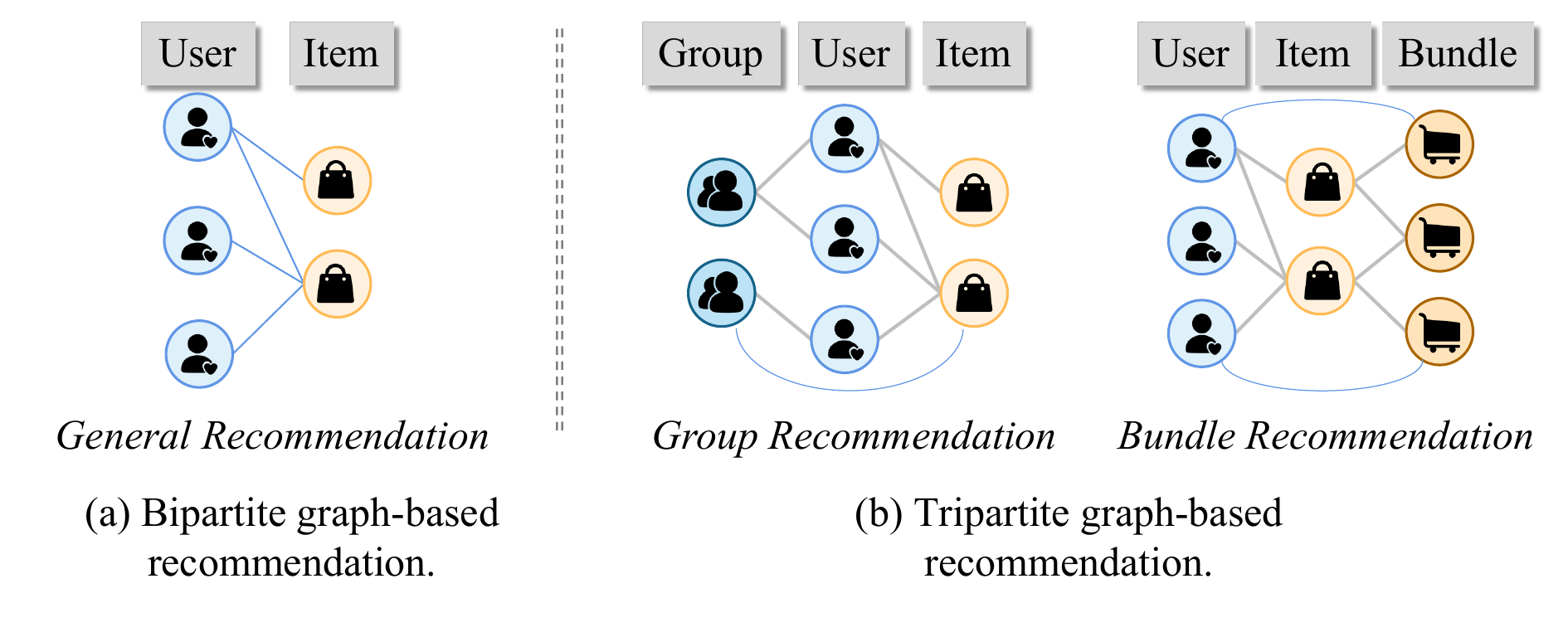}

    \caption{Comparison of bipartite and tripartite graph-based recommendation.}
    \label{fig:bipartite-tripartite}
    \vspace{-1em}
\end{figure}

Tripartite graph-based recommendations differ substantially from general recommender systems. For instance, group recommendations suggest items to diverse-interest user groups, e.g., recommending a travel route to a group of tourists, and bundle recommendations propose multiple items to a user simultaneously.
Unlike bipartite graphs that encompass interactions between two entities (users and items), tripartite graphs engage three types of entities: users, items, and groups/bundles, allowing arbitrary groupings of users and items. 
Despite the relative abundance of user-item interactions, interactions between recommended objects (user/group) and recommendees (bundle/item) are markedly sparse \cite{hao2023multi, hao2021pre, zhu2022variational} (demonstrated in Fig. \ref{fig:bipartite-tripartite}). 
The complexities and sparsity, particularly noticeable in scenarios like group recommendations where groups engage with fewer items, inhibit the efficacy of standard pair-wise ranking methods during positive and negative sampling. This exacerbates the difficulty of accurately modeling group nodes' latent semantics in a tripartite graph and, subsequently, making precise recommendations.

Mitigation of data sparsity has been critically addressed through various sophisticated mechanisms in group and bundle recommendation models. Attention mechanisms manage data sparsity and cold-start issues in group or bundle recommendations by dynamically prioritizing interactions and aggregating item significance \cite{guo2020group, cao2018attentive, 10.1145/3534678.3539229}. Multi-task learning is also an effective way, which enhances robustness in user-group and user-item interactions by jointly optimizing parallel tasks and effectively utilizing sparse data through integrating various interaction types \cite{jia2021hypergraph, chen2019matching}. Additionally, graph-based approaches navigate data sparsity by traversing user-item, user-group, and user-bundle interactions within a heterogeneous graph, facilitating information propagation and ensuring cohesive embedding of item, group, and bundle semantics \cite{9546546}. Thus, attention application, multi-task learning, and graph-based propagation, skillfully interweave varied interactions and affiliations to adeptly mitigate challenges in sparse data scenarios. However, these methods face challenges in extreme cold-start scenarios due to their reliance on a minimum amount of positive and negative samples for pair-wise ranking training in the recommendation-related loss. Details can be seen in Appendix \ref{ap:rela}.

Addressing challenges in tripartite graph-based recommendations involves two core issues. First, the substantial sparsity in interactions between recommended objects and recommendees necessitates novel metrics to bridge the semantic gap between lower user-item interactions and user-group/item-bundle affiliations and higher relationships between recommended objects and recommendees. Second, the effective use of metrics as proxy supervision signals becomes critical to guide the training of the recommendation model, ensuring refined embeddings for recommended objects and recommendees with rare interactions, which facilitates accurate future recommendations.

To address these challenges, we propose a \underline{C}onsistency and \underline{D}iscrepancy-based contrastive learning \underline{R}ecommendation method, i.e., CDR, for tripartite graph-based recommendation. We initiate with two new metrics, consistency and discrepancy, between the recommended object and recommendee nodes in the tripartite graph, defining a hard multi-objective optimization problem to establish high-order relationships based on abundant user-item interactions and affiliations. We define reachable and non-reachable meta-path between recommended objects and recommendees, mediated/blocked by other nodes in the tripartite graph. We efficiently pre-calculate two coefficients related to the solution, encouraging similar (consistency) or dissimilar (discrepancy) embeddings of recommended objects and recommendees, utilizing graph convolutional networks (GCN) limit theory.

Subsequently, a contrastive inspired CD loss, integrating the designed consistency and discrepancy metrics as positive and contrastive supervision signals, is introduced to train the model without requiring direct interactions between objects and recommendees. Furthermore, when direct interactions become available, they can fine-tune consistency and discrepancy metrics, enhancing recommendation accuracy.
The contributions of this paper can be concretely summarized into four folds as follows:
\begin{itemize}[leftmargin=0.5cm]
\item We propose a new contrastive graph learning-inspired tripartite graph-based recommender system, i.e., CDR, based on new supervision signals and a new optimization procedure. CDR is shown to be robust to the extreme cold-start case where no direct interaction exists between the recommended objects and the recommendees. 
\item Two novel metrics, i.e., consistency and discrepancy, are proposed by solving a multi-objective optimization problem based on reachable/non-reachable meta-paths between recommended objects and recommendees, with available user-item interactions
and user-group/item-bundle affiliations. The metrics can be efficiently pre-calculated before optimization with GCN limit theory. 
\item Furthermore, a contrastive learning-inspired novel CD loss is proposed accordingly, where the introduced consistency and discrepancy metrics can be seamlessly integrated into the contrastive objective as the positive and contrastive supervision signals to learn node representations, without the requirement of available direct user-bundle/group-item interactions. 
\item Extensive experiments on real-world datasets demonstrate the superiority of the proposed CDR model, where the advantages of the proposed consistency and discrepancy metrics and the contrastive learning-based CD loss are clearly demonstrated by the ablation study.
\end{itemize}

\section{Preliminaries}
\label{section:preliminary}
In this section, we introduce the symbols used in this paper, define related concepts, and formally formulate the problem of tripartite graph-based recommendation as the preliminary knowledge of the proposed CDR model.

\begin{table}[]
\centering
\caption[]{Summary of notations.}
\label{tab:notation}
\resizebox{0.4\textwidth}{!}{
\begin{tabular}{c||l}
\hline
\textbf{Notation} & \textbf{Descriptions} \\ \hline \hline
$\mathcal{T}$, $\mathcal{O}$, $\mathcal{M}$ & The sets of tuple, object, and member nodes\\
$|\mathcal{T}|$, $|\mathcal{O}|$, $|\mathcal{M}|$ & The number of tuple, object, and member nodes\\
$t$, $o$, $m$ & A tuple / object / member node\\
($v_1$, $v_2$) & A node pair\\
$\boldsymbol{Y}$, $\boldsymbol{X}$, $\boldsymbol{Z}$ & \begin{tabular}[c]{@{}l@{}} The matrices of tuple interactions, member \\interactions, and tuple-member affiliations \end{tabular}\\
$e_{t}$, $e_{o}$, $e_{m}$ & The embedding of tuple, object, and member nodes\\
$\mathcal{N}(t)$, $\mathcal{N}(o)$, $\mathcal{N}(m)$ & \begin{tabular}[c]{@{}l@{}} The set of neighbors of tuple node $t$, object node $o$, \\and member node $m$ \end{tabular}\\
$deg_{v}$, $\delta_{v_1, v_2}$ & \begin{tabular}[c]{@{}l@{}} The degree of node $v$ and the composite term of the \\degrees of nodes $v_1$ and $v_2$\end{tabular}\\
$c_{v_1,v_2}$, $d_{v_1,v_2}$ & The consistency and discrepancy metrics \\
$\boldsymbol{C}^{\Phi}$, $\boldsymbol{D}^{\Phi}$ & \begin{tabular}[c]{@{}l@{}} The consistency and discrepancy sub-matrices extracted \\from meta-path $\Phi$ \end{tabular}\\
$\boldsymbol{C}_{tuple}$, $\boldsymbol{D}_{tuple}$ & \begin{tabular}[c]{@{}l@{}} The consistency and discrepancy matrices extracted \\from tuple interactions  \end{tabular}\\
$\boldsymbol{C}_{member}$, $\boldsymbol{D}_{member}$ & \begin{tabular}[c]{@{}l@{}} The consistency and discrepancy matrices extracted \\from member interactions and tuple-member\\ affiliations\end{tabular}\\ \hline
\end{tabular}
}
\end{table}

\subsection{Relevant Concepts}
\label{subsec:concept}
\begin{myDef}
\textbf{Tripartite Graph} \cite{sun2013mining}. 
Given three node sets $\mathcal{T}$, $\mathcal{O}$, and $\mathcal{M}$, $\mathcal{V} = \mathcal{T} \cup \mathcal{O} \cup \mathcal{M}$, a graph $G=(\mathcal{V},\mathcal{E}$) is called a tripartite graph, if the edge set $\mathcal{E} \subseteq \mathcal{T} \times \mathcal{O} \cup \mathcal{O} \times \mathcal{M} \cup \mathcal{T} \times \mathcal{M}$, where $\times$ represents Cartesian product. In tripartite graph-based recommendation, $G$ is a heterogeneous graph since $\mathcal{T}$, $\mathcal{O}$, and $\mathcal{M}$ represent three different types of nodes, i.e., user, item, and user/item groups.
\end{myDef}

\begin{myDef}
\textbf{Tripartite Graph-Based Recommendation}.
The tripartite graph-based recommendation contains three types of entities, i.e., user, item, and user/item group (where item groups are known as bundles in the literature). To simplify the discussion, we use the term \textbf{\underline{tuples}} to represent user groups and item bundles, use the term \textbf{\underline{members}} to represent users in group recommendation and items in bundle recommendation, and use the term \textbf{\underline{objects}} to represent items in group recommendation and users in bundle recommendation. For simplicity, we use tuple interactions to represent the interaction between tuple and object, and member interactions to represent interactions between member and object.
\end{myDef}
Tripartite graph-based recommendation unifies group and bundle recommendations as the interaction prediction problem between tuples and objects, where the proposed CDR can be viewed as a universal solution.

\begin{myDef}
\textbf{Meta-Path} \cite{sun2011pathsim}.
A (reachable) meta-path $\Phi$ defined on the heterogeneous graph $G=(\mathcal{V}, \mathcal{E})$ is a path pattern in the form of $V_{1} - V_{2} -\cdots -V_{l+1}$, where $V_{k} \in \{\mathcal{T}, \mathcal{O}, \mathcal{M}\}$. Specifically, $\Phi$ describes the composite relationship $R=R_{V_{1} V_{2}} \circ R_{V_{2} V_{3}} \circ \cdots \circ R_{V_{l} V_{l+1}}$ between the head node $V_{1}$ and the tail node $V_{l+1}$, where $\circ$ denotes the composition operator and $R_{V_{k} V_{k+1}}$ denotes the pairwise relationship defined in the edge set $\mathcal{E}$.
\end{myDef}
 
\begin{figure}[htbp]
\centering
\subfigure[Reachable meta-path Tuple-Member-Tuple.]
{
\includegraphics[width=0.38\linewidth]{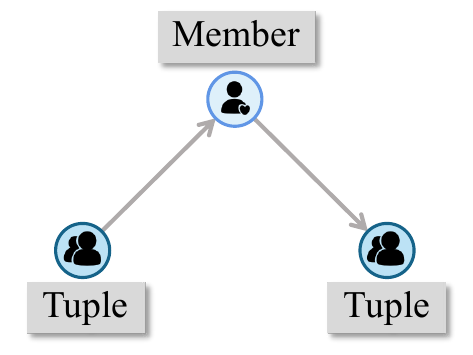}
\label{fig:subfig1}
}
\hspace{1em}
\subfigure[Non-reachable meta-path Tuple$\sim$Member-Tuple.]
{
\includegraphics[width=0.38\linewidth]{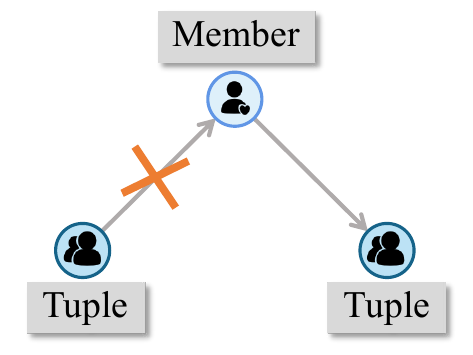}
\label{fig:subfig2}
}
\caption{Two kinds of meta-paths, as examples, for deriving consistency and discrepancy metrics.}
\label{fig:metapath}
\end{figure}

We extend the meta-path schema definition to include \textbf{non-reachable information} between nodes, allowing explicit consideration of negative pairwise relationships. Specifically, Tuple$\sim$ Member$-$Tuple represents non-reachable meta-path schemas, with "$-$" signifying an edge and "$\sim$" no edge between nodes, as depicted in Fig. \ref{fig:metapath}.

\subsection{Problem Formulation}

Consider the sets of tuple nodes $\mathcal{T} = \left \{ t_{1},\ldots,t_{|\mathcal{T}|}\right \}$, object nodes $\mathcal{O} = \left \{ o_{1},\ldots,o_{|\mathcal{O}|}\right \}$, and member nodes $\mathcal{M} = \left \{ m_{1},\ldots,m_{|\mathcal{M}|}\right \}$, where $|\mathcal{T}|$, $|\mathcal{O}|$, and $|\mathcal{M}|$ represent the number of the tuple, object, and member nodes, respectively. Three relations could exist among $\mathcal{T}$, $\mathcal{O}$, and $\mathcal{M}$, i.e., tuple interactions, member interactions, and member-tuple affiliations. We use a sparse matrix $\boldsymbol{Y} \in\{0,1\}^{|\mathcal{T}| \times |\mathcal{O}|}$ to represent tuple interactions, where $y_{to} = 1$ indicates that the tuple node $t$ has interacted with the object $o$, and $y_{to} = 0$ indicates otherwise. Since members can be arbitrarily grouped into tuples, $\boldsymbol{Y}$ can be extremely sparse, and in the extreme cold-start case, $\boldsymbol{Y}$ can be an empty matrix. Similarly, member interactions are represented by the sparse matrix $\boldsymbol{X} \in\{0,1\}^{|\mathcal{M}| \times |\mathcal{O}|}$, where $x_{mo} = 1$ indicates that the member $m$ has interacted with the object $o$, and $x_{mo}$ = 0 indicates otherwise. $\boldsymbol{X}$ is usually much denser than $\boldsymbol{Y}$ due to the comparatively abundant member interactions. Finally, we denote the observed affiliations as $\boldsymbol{Z} \in\{0,1\}^{|\mathcal{T}| \times |\mathcal{M}|}$, where the element $z_{tm}=1$ means the member $m$ is affiliated to the tuple $t$, and $z_{tm}=0$ otherwise.

The aim of tripartite graph-based recommendation is to generate a top-K list of objects (bundles/items) for the recommendee (users/user groups) using member interactions $\boldsymbol{X}$, tuple interactions $\boldsymbol{Y}$, and tuple-member affiliations $\boldsymbol{Z}$. It also addresses the \textbf{extreme cold-start} where no tuple interactions (i.e., $\boldsymbol{Y}$ is a zero-matrix) exist. Table \ref{tab:notation} summarizes symbols.

\section{Methodology}
\label{section:methodology}

\subsection{Tuple-Object Metric Learning}
\label{section:consistency and discrepancy}

In the first stage of the proposed CDR algorithm, we compute consistency and discrepancy metrics using defined reachable/non-reachable meta-paths and abundant member interactions plus member-tuple affiliations. These metrics gauge the similarity or difference between pairs of recommended object and recommendee nodes, crucial for the positive and negative relations needed by the contrastive learning-inspired CD loss, discussed subsequently.
\subsubsection{Optimization Objective}

The derivation of consistency and discrepancy metrics begins with an auxiliary graph embedding learning task, intending to map all nodes of the tuple-member-object tripartite graph into an embedding space while maximally preserving tuple-member affiliation and object-member interaction information in the node embeddings. Specifically, utilizing the positive and negative pairwise relationships between tuples/members and members/objects, we aim to optimize the following objectives:
\begin{equation}
\label{equation:tuple_member_maxmin}
\max \sum_{m \in \mathcal{N}(t)} e_{t}^{\top} e_{m}, \quad
\min \sum_{\tilde{m} \notin \mathcal{N}(t)} e_{t}^{\top} e_{\tilde{m}},
\end{equation}
\begin{equation}
\label{equation:object_member_maxmin}
\max \sum_{m \in \mathcal{N}(o)} e_{o}^{\top} e_{m}, \quad
\min \sum_{\tilde{m} \notin \mathcal{N}(o)} e_{o}^{\top} e_{\tilde{m}},
\end{equation}
where $e$ is the node embedding, $\mathcal{N}(t)$ and $\mathcal{N}(o)$ are the neighbor sets of tuple $t$ and object $o$, and $m$ and $\tilde{m}$ are the first-order member neighbors and non-first-order neighbor nodes (members) of tuple $t$ and object $o$, respectively. To simplify the discussion, we only consider nodes $t, m, \tilde{m}, o$ in the objective, but the generalization of Eqs. (\ref{equation:tuple_member_maxmin}) and (\ref{equation:object_member_maxmin}) to include all nodes in the graph is straightforward.

Prior research primarily optimizes object and member embeddings (user-item interactions) as per Eq. (\ref{equation:object_member_maxmin}), often using techniques like attention mechanisms \cite{vaswani2017attention} to aggregate member embeddings into tuple embeddings, neglecting Eq. (\ref{equation:tuple_member_maxmin}). Such simple aggregation from member to tuple level lacks theoretical and explanatory backing. Contrarily, our method effectively optimizes all four objectives in Eqs. (\ref{equation:tuple_member_maxmin}) and (\ref{equation:object_member_maxmin}) concurrently.

\subsubsection{Message Passing Mechanism}

Optimizing Eq. (\ref{equation:object_member_maxmin}) involves message passing among heterogeneous nodes. Utilizing the LightGCN strategy, celebrated for effectively mining bipartite graph pair-wise relations \cite{he2020lightgcn}, we employ it as our message-passing mechanism. The user embedding information aggregation process in the LightGCN network, incorporating self-connection, is formalized as follows:
\begin{equation}
\label{equation:LightGCN}
\begin{split}
    e_{m}^{(l+1)}=\frac{1}{deg_{m}+1} e_{m}^{(l)}&+\sum_{t \in \mathcal{N}(m)} \frac{1}{\sqrt{deg_{m}+1} \sqrt{deg_{t}+1}}e_{t}^{(l)}\\&+\sum_{o\in \mathcal{N}(m)} \frac{1}{\sqrt{deg_{m}+1} \sqrt{deg_{o}+1}} e_{o}^{(l)},
\end{split}
\end{equation}
where $e^{(l)}$ represents the node embedding at layer $l$, and $deg_{v}$ represents the degree of node $v$, respectively.

\subsubsection{Solution Based on the Limit Theory of GCN}

The direct generalization of Eq. (\ref{equation:LightGCN}) to solve Eqs. (\ref{equation:tuple_member_maxmin}) and (\ref{equation:object_member_maxmin}) is computationally expensive due to multiple aggregation objectives. Moreover, it is difficult to decide how many aggregation processes are needed to acquire a good embedding conducive to recommendation. Inspired by UltraGCN\cite{mao2021ultragcn}, we note that a convergent form of $e_{m}$ can be obtained with infinite propagation layers. Specifically, according to the aggregation formula of graph convolution, after passing through infinite layers of aggregation, the embeddings of the model will converge to the following form: 
\begin{equation}
\label{eq:limit_emb}
e_{m}=\lim _{l \rightarrow \infty} e_{m}^{(l+1)}=\lim _{l \rightarrow \infty} e_{m}^{(l)},
\end{equation}
where the node embeddings are consistent before and after the convolution in the limit. Based  on Eq. (\ref{eq:limit_emb}), the message passing mechanism defined in Eq. (\ref{equation:LightGCN}) can be simplified as:
\begin{equation}
\label{equation:em}
\begin{aligned}
e_{m}=\sum_{t \in \mathcal{N}\left(m\right)}
\frac{1}{deg_{m}} \sqrt{\frac{deg_{m}+1}{deg_{t}+1}} e_{t} + \sum_{o \in \mathcal{N}\left(m\right)} 
\frac{1}{deg_{m}} \sqrt{\frac{deg_{m}+1}{deg_{o}+1}} e_{o}. 
\end{aligned}
\end{equation}
We denote the composite term of the degree of member $m$ and tuple $t$ (object $o$) as $\delta_{m, t}$ ($\delta_{m, o}$) defined as follows:
\begin{equation}
\label{equation:delta}
\delta_{m, t}=\frac{1}{deg_{m}} \sqrt{\frac{deg_{m}+1}{deg_{t}+1}}, \quad \delta_{m, o}=\frac{1}{deg_{m}} \sqrt{\frac{deg_{m}+1}{deg_{o}+1}},
\end{equation}
Eq. (\ref{equation:em}) can be simplified into the following form:
\begin{equation}
\label{equation:e_m}
e_{m}=\sum_{t \in \mathcal{N}\left(m\right)} \delta_{m, t} e_{t}+\sum_{o \in \mathcal{N}\left(m\right)} \delta_{m, o} e_{o}.
\end{equation}
Essentially, Eq. (\ref{equation:e_m}) uses tuple embeddings and object embeddings to represent member embeddings. To further utilize tuple-member affiliations, we can substitute Eq. (\ref{equation:e_m}) into Eq. (\ref{equation:tuple_member_maxmin}) and maximize the two parts simultaneously to approximate the final result, where the objective can be re-formulated as:
\begin{equation}
\label{equation:tm_max}
\begin{split}
    \max \sum_{m \in \mathcal{N}(t)} e_{t}^{\top} e_{m}\simeq &\max \sum_{m \in \mathcal{N}(t)} \sum_{t^{\prime} \in \mathcal{N}\left(m\right)} \delta_{m, t^{\prime}} e_{t}^{\top} e_{t^{\prime}}+\\
    &\max \sum_{m \in \mathcal{N}(t)} \sum_{o \in \mathcal{N}\left(m\right)} \delta_{m, o} e_{t}^{\top} e_{o}.
\end{split}
\end{equation}
Since the member embedding $e_m^\prime$ is eliminated from the righ-hand side of Eq. (\ref{equation:tm_max}), we can utilize the relationship of member nodes without training their embeddings.

\subsubsection{Solution with Reachable/Non-Reachable Meta-Paths}

While Eq. (\ref{equation:tm_max}) is notably simplified, it still demands gradient-based optimization. The optimization is further streamlined by introducing reachable and non-reachable meta-paths from Section \ref{subsec:concept}, allowing the maximization portion of Eq. (\ref{equation:tm_max}) to be redefined as follows:
\begin{equation}
\label{equation:loss_tm_max}
\begin{split}
    \mathcal{L}_{tm\_max }=&\sum_{\left(t, m, t^{\prime}\right) \in Q_{t m t^\prime}} \delta_{m, t^{\prime}}\left(1-\sigma\left(e_{t}^{\top} e_{t^{\prime}}\right)\right)\\+&\sum_{\left(t, m, o\right) \in Q_{t m o}} \delta_{m, o}\left(1-\sigma\left(e_{t}^{\top} e_{o}\right)\right),
\end{split}
\end{equation}
where $\sigma(\cdot)$ is the sigmoid function, $Q_{tmt^\prime}$ is the set of reachable Tuple-Member-Tuple meta-paths ($t,m,t^{\prime}$) composed of tuple node $t$ and its first-order member neighbors $m$ and second-order tuple neighbors $t^{\prime}$. Similarly, $Q_{tmo}$ is the set of reachable Tuple-Member-Object meta-paths ($t, m, o$): 
\begin{align}
Q_{t m t^\prime}=\left\{\left(t, m, t^{\prime}\right) \mid m \in \mathcal{N}(t), t^{\prime} \in \mathcal{N}\left(m\right)\right\}, \notag \\
Q_{t m o}=\left\{\left(t, m, o\right) \mid m \in \mathcal{N}(t), o \in \mathcal{N}\left(m\right)\right\}.
\end{align}
The minimization part of Eq. (\ref{equation:tuple_member_maxmin}) can be reformulated as:
\begin{equation}
\begin{split}
    \mathcal{L}_{tm\_min}=\sum_{\left(t, \tilde{m}, t^{\prime}\right) \in Q_{t \tilde{m} t^\prime}} \delta_{\tilde{m}, t^{\prime}}\sigma\left(e_{t}^{\top} e_{t^{\prime}}\right) +\sum_{\left(t, \tilde{m}, o\right) \in Q_{t \tilde{m} o}} \delta_{\tilde{m}, o}\sigma\left(e_{t}^{\top} e_{o}\right).
\end{split}
\label{equation:loss_tm_min}
\end{equation}

The set $Q_{t \tilde{m} t^\prime}$ diverges from $Q_{t m t^\prime}$ by representing non-reachable Tuple$\sim$Member-Tuple meta-paths. Here, triples $(t, \tilde{m}, t^\prime)$ in $Q_{t \tilde{m} t^\prime}$ denote member $\tilde{m}$ is not adjacent to tuple $t$, but is to tuple $t^\prime$. Additionally, $Q_{t \tilde{m} o}$ defines non-reachable Tuple$\sim$Member-Object meta-paths as follows:
\begin{align}
Q_{t \tilde{m} t^\prime}=\left\{\left(t, \tilde{m}, t^{\prime}\right) \mid \tilde{m} \notin \mathcal{N}(t), t^\prime \in \mathcal{N}\left(\tilde{m}\right)\right\}, \notag \\
Q_{t \tilde{m} o}=\left\{\left(t, \tilde{m}, o\right) \mid \tilde{m} \notin \mathcal{N}(t), o \in \mathcal{N}\left(\tilde{m}\right)\right\}.
\end{align}

\subsubsection{Final Objective}

Examining the loss functions $\mathcal{L}_{tm\_max}$ and $\mathcal{L}_{tm\_min}$ from Eq. (\ref{equation:loss_tm_max}) and Eq. (\ref{equation:loss_tm_min}), the initial terms engage with $t$ and $t^{\prime}$, while the secondary involve $t$ and $o$. Losses for pairs $t, t^{\prime}$ and $t, o$ are structured accordingly. Focusing on pair ($t, t^{\prime}$), the losses in $\mathcal{L}_{tm\_max}$ and $\mathcal{L}_{tm\_min}$ integrate as follows:
\begin{equation}
\begin{split}
    l_{t, t^{\prime}}=&\sum_{m \in \mathcal{N}(t), m \in \mathcal{N}\left(t^{\prime}\right)} \delta_{m, t^{\prime}}\left(1-\sigma\left(e_{t}^{\top} e_{t^{\prime}}\right)\right)\\+&\sum_{\tilde{m}  \notin \mathcal{N}(t), \tilde{m} \in \mathcal{N}\left(t^\prime\right)} \delta_{\tilde{m}, t^\prime} \sigma\left(e_{t}^{\top} e_{t^\prime}\right).
\end{split}
\end{equation}
\noindent We omit the constant that is independent to $e_t$ and $e_{t^{\prime}}$, and use $c_{t,t^{\prime}}$ and $d_{t,t^{\prime}}$ to denote the constants that concern only $e_t$ and $e_{t^{\prime}}$, where
$c_{t, t^{\prime}}=\sum_{m \in \mathcal{N}(t), m \in \mathcal{N}\left(t^{\prime}\right)} \delta_{m, t^{\prime}}, \quad d_{t, t^{\prime}}=\sum_{\tilde{m}  \notin \mathcal{N}(t), \tilde{m} \in \mathcal{N}\left(t^{\prime}\right)} \delta_{\tilde{m}, t^{\prime}}$, the loss function of a node pair ($t, t^{\prime}$) can be rewritten as follows:
\begin{equation}
l_{t, t^{\prime}}=\left(d_{t, t^{\prime}}-c_{t, t^{\prime}}\right) \sigma\left(e_{t}^{\top} e_{t^{\prime}}\right). \label{equation:l_tt}
\end{equation}
All node pairs of arbitrary type can derive their consistencies $c_{vi,vj}$ and discrepancies $d_{vi,vj}$ by predefined reachable and non-reachable meta-paths. For an arbitrary node pair ($t,o$), we construct meta-path sets $Q_{tmo}$ and $Q_{t\tilde{m} o}$ subsequently deriving its loss function with similar steps applied to other node pair types:
\begin{equation}
\label{eq:to}
    l_{t,o} = (d_{t,o} - c_{t,o})\sigma(e_t^{\top}e_{o}).
\end{equation}

Calculating $c_{t, o}$ and $d_{t, o}$ parallels $c_{t, t^{\prime}}$ and $d_{t, t^{\prime}}$. Given the asymmetry in calculations like Eq. (\ref{equation:l_tt}) and Eq. (\ref{eq:to}) regarding node pair order, and possible pair types (tuple-tuple, tuple-object, object-tuple, and object-object), coefficients such as $c_{t, o}$ and $d_{t, o}$ are also computed. The loss function for arbitrary type node pair ($v_1,v_2$) using meta-paths with user nodes as midpoints is:
\begin{equation}
\label{equation:combine_loss}
   \mathcal{L}=\sum_{v_1, v_2 \in \mathcal{T\cup O}}\left(d_{v_1,v_2}-c_{v_1,v_2}\right) \sigma\left(e_{v_1}^{\top} e_{v_2}\right),
\end{equation} where coefficients $c_{v_1, v_2}$ and $d_{v_1, v_2}$ can be pre-calculated as follows:
\begin{equation}
\label{equation:c_d}
\begin{split}
    c_{v_1, v_2}&=\sum_{m \in \mathcal{N}(v_1), m \in \mathcal{N}\left(v_2\right)} \delta_{m,v_2}, \\
    d_{v_1, v_2}&=\sum_{\tilde{m}  \notin \mathcal{N}(v_1), \tilde{m} \in \mathcal{N}\left(v_2\right)} \delta_{\tilde{m},v_2}.
\end{split}
\end{equation}
\subsubsection{Consistency and Discrepancy Metrics}
\label{subsec:Sim}

In general recommendations, coefficients $c_{v_1,v_2}$ and $d_{v_1,v_2}$ in Eq. (\ref{equation:combine_loss}) determine the embedding distance between recommended object and recommendee pairs, with $c_{v_1,v_2}$ and $d_{v_1,v_2}$ respectively encouraging closer and more distant embeddings, termed as \textbf{consistency} and \textbf{discrepancy}. For $v_{1}, v_{2} \in \mathcal{T}\cup \mathcal{O}$,  \textbf{tuple-object relationships} derive from $c_{v_1,v_2}$ and $d_{v_1,v_2}$ without needing direct interactions, seamlessly integrating with contrastive learning by providing positive and contrastive supervision signals.

\subsection{Contrastive Learning-Inspired CD Loss}

Utilizing consistency and discrepancy from subsection \ref{subsec:Sim}, we introduce the contrastive learning-inspired CD loss here, employing them as positive and negative supervision signals to formulate the contrastive objective.

\subsubsection{Motivation of the CD Loss}
\label{sec:motivation}

Given Eq. (\ref{equation:combine_loss}), a direct model training strategy might use a general recommendation algorithm with $d_{v_1,v_2}-c_{v_1,v_2}$ as supervision signals. However, in tripartite graph-based recommendations, the subtraction ($d_{v_1,v_2}-c_{v_1,v_2}$) potentially erases unique consistency and discrepancy information. For example, cases (dv1,v2=1, cv1,v2=0.99) and (dv1,v2=0.01, cv1,v2=0) both result in a subtraction value of 0.01, despite significant differences in absolute values. Unlike bipartite recommendations, where subtractions $d-c$ are clearly 1 or -1, we propose a new contrastive learning-based objective using consistency and discrepancy independently as positive and contrastive supervision signals.

\subsubsection{CD Loss}
To further utilize the consistency and discrepancy, inspired by contrastive learning\cite{gutmann2010noise}, we design a CD loss for tripartite graph-based recommendation, leveraging consistency and discrepancy for any node pair ($v_1,v_2$):
\begin{equation}
\mathcal{L}=\sum_{(v_1, v_2) \in Q}-\log \frac{c_{v_1, v_2} \exp \left(\cos \left(\mathbf{e}_{v_1}, \mathbf{e}_{v_2}\right) / \tau\right)}{\sum_{\tilde{v}_2 \in \mathcal{T\cup O}} d_{v_1, \tilde{v}_2} \exp \left(\cos \left(\mathbf{e}_{v_1}, \mathbf{e}_{\tilde{v}_2}\right) / \tau\right)},
\label{equation:contrastive}
\end{equation}
where $Q=\{(v_1, v_2) \mid c_{v_1,v_2}>0\quad and\quad v_1,v_2 \in  \mathcal{T\cup O}\}$ is the set of node pairs whose consistency is positive, and $\cos \left(\cdot, \cdot \right)$ is the cosine similarity measure. In Eq. (\ref{equation:contrastive}), the consistency and discrepancy metrics are independently used to select positive and negative supervision signals. During the optimization process, the consistency $c_{v_1,v_2}$ encourages the embeddings of nodes $v_1$ and $v_2$ to be 
close, while the discrepancy $d_{v_1 ,\tilde{v}_2}$ constrains the embeddings to be distant with each other.

The contrastive loss in Eq. (\ref{equation:contrastive}) introduces a temperature parameter $\tau$, adjustable based on dataset noise levels; a larger $\tau$ suits higher noise, focusing model updates on easily distinguishable nodes, while a smaller $\tau$ is used for higher-quality data to strictly differentiate similar nodes.

\begin{figure*}[htbp]
	\centering
	\includegraphics[width=0.85\textwidth,height=0.28\textheight]{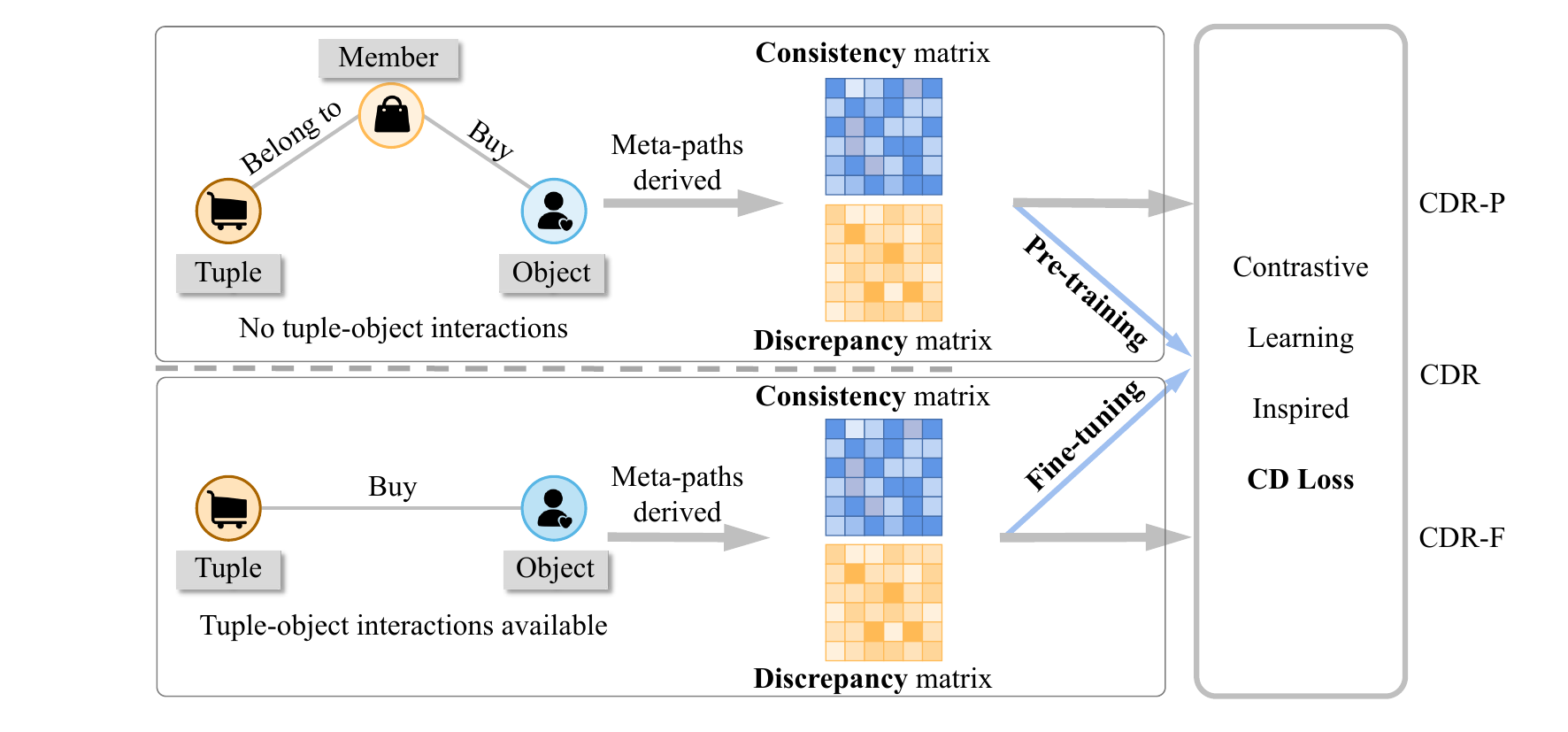}
    \caption{CDR Model Structure. Upper pre-training with tuple-member affiliations and member interactions, lower fine-tuning with tuple interactions, featuring variants CDR-P and CDR-F. CDR-P is for extreme cold-start scenes, whereas CDR-F is a variant that only uses tuple interactions.}
	\label{fig:tuple_recommendation}
\end{figure*}

\subsection{The Overall Procedure of the CDR}
Fig. \ref{fig:tuple_recommendation} depicts the CDR structure, which computes consistency and discrepancy metrics between tuples and objects either from member interactions and tuple-member affiliations (upper part) or solely from tuple interactions (lower part). The former method, referred to as the \textbf{pre-training stage}, utilizes member interactions, while the latter can \textbf{fine-tune} metrics using available tuple interactions, aiding the extraction of node representations.

\subsubsection{Two Variants of CDR}
This subsection details the pre-training and fine-tuning of CDR via two variants: CDR-P and CDR-F. CDR-P, indicating pre-training, is trained using metrics from member interactions and tuple-member affiliations, applicable in extreme cold-start scenarios without tuple interactions. Conversely, CDR-F (fine-tuning) is trained using tuple interaction metrics, aligning closer with traditional recommendation models. The full CDR model is formulated by integrating both CDR-P and CDR-F variants.

We denote the embedding matrices for the pre-training and fine-tuning stages as $\mathbf{E}^{p}$ and $\mathbf{E}^{f}$, respectively. The matrix $\mathbf{E}^{p}$ initializes $\mathbf{E}^{f}$ in the fine-tuning phase, providing improved initial embedding values and fostering quicker model convergence. We also retain $\mathbf{E}^{p}$ as a static feature and concatenate it with $\mathbf{E}^{f}$ to establish the final embedding $\mathbf{E}$:

\begin{equation}
\mathbf{E} = \mathbf{E}^{f} \parallel \mathbf{E}^{p},
\end{equation}

where $\parallel$ is the concatenate operation. $\mathbf{E}$ is used for subsequent fine-tuning and recommendation. Our model only updates $\mathbf{E}^{f}$ but not $\mathbf{E}^{p}$ in the fine-tuning stage, which can avoid the node preferences extracted in the pre-training stage being erased by downstream tasks, where member interaction and member-tuple affiliation information is substantially preserved. 

\subsubsection{Generalization of Consistency and Discrepancy}

We seek to extend the consistency and discrepancy metrics from Section \ref{subsec:Sim} for the fine-tuning stage. During pretraining, eight meta-path types are utilized. Specifically, $TMT$, $TMO$, $OMT$, and $OMO$ stand for node types on meta-paths, and each item can represent both reachable and non-reachable meta-paths with the same node type. Matrices, subscripted with \textit{member}, denote consistency and discrepancy across these meta-paths:
\begin{equation}
\begin{split}
\boldsymbol{C}_{member}&=\left[\begin{array}{ll}
\boldsymbol{C}^{TMT} & \boldsymbol{C}^{TMO} \\
\boldsymbol{C}^{OMT} & \boldsymbol{C}^{OMO}
\end{array}\right], \\
\boldsymbol{D}_{member}&=\left[\begin{array}{ll}
\boldsymbol{D}^{TMT} & \boldsymbol{D}^{TMO} \\
\boldsymbol{D}^{OMT} & \boldsymbol{D}^{OMO}
\end{array}\right],
\end{split}
\end{equation}
where $\boldsymbol{C}^{TMT}=\left[c_{t,t^\prime}\right]^{|\mathcal{T}| \times |\mathcal{T}|}$ represents the consistency extracted through the meta-path Tuple-Member-Tuple ($TMT$), as in the other sub-matrices. Elements in these matrices are used in the pre-training stage.

In the fine-tuning stage, with accessible tuple interactions, additional four types of meta-paths are considered, namely Tuple-Object-Tuple ($TOT$) and Object-Tuple-Object ($OTO$) for consistency, Tuple$\sim$Object-Tuple ($TOT$), and Object$\sim$Tuple-Object ($OTO$) for the discrepancy, alongside direct $TO$ and $OT$ relationships, given the asymmetry of consistency and discrepancy. Corresponding matrices are denoted as follows:
\begin{equation}
\begin{split}
\boldsymbol{C}_{tuple}&=\left[\begin{array}{ll}
\boldsymbol{C}^{TOT} & \boldsymbol{C}^{TO} \\
\boldsymbol{C}^{OT} & \boldsymbol{C}^{OTO}
\end{array}\right], \\
\boldsymbol{D}_{tuple}&=\left[\begin{array}{ll}
\boldsymbol{D}^{TOT} & \boldsymbol{D}^{TO} \\
\boldsymbol{D}^{OT} & \boldsymbol{D}^{OTO}
\end{array}\right].
\end{split}
\end{equation}
Given that one-hop meta-paths $TO$ and $OT$ cannot be utilized directly via Eq. (\ref{equation:c_d}), a redefinition is required. Drawing inspiration from \cite{mao2021ultragcn}, we propose a generalized form for consistency and discrepancy by modifying $\delta$ in Eq. (\ref{equation:c_d}) as follows:
\begin{equation}
\boldsymbol{C}^{TO}=\left[\delta_{to} * Y_{to} \right]^{|\mathcal{T}| \times |\mathcal{O}|}, \quad
\boldsymbol{D}^{TO}=\left[\delta_{to} \right]^{|\mathcal{T}| \times |\mathcal{O}|},
\end{equation}
where $\boldsymbol{Y} \in\{0,1\}^{|\mathcal{T}| \times |\mathcal{O}|}$ is the tuple interaction matrix. The node relationship on the meta-path $OT$ is similar to $TO$. With the elements in $\boldsymbol{C}_{tuple}$ and $\boldsymbol{D}_{tuple}$, the model can be fine-tuned to get a better recommendation performance. The algorithm procedure of CDR is presented in Appendix \ref{ap:algo}.

\section{Experiments}
\label{section:experiment}

In this section, we detail extensive experiments on various real-world datasets to validate CDR's effectiveness, addressing the following four research questions:

\begin{itemize}[leftmargin=0.5cm]
    \item \textbf{RQ1} How does CDR compare to baselines in sparse tuple interactions?
    \item \textbf{RQ2} How does CDR compare to other methods in extreme cold-start scenarios with no tuple interactions?
    \item \textbf{RQ3} How do CDR's components benefit its performance?
    \item \textbf{RQ4} How does the hyperparameter setting of temperature impact CDR's performance? 
\end{itemize}

Furthermore, we conduct extra analysis to compare the recommendation diversity between CDR and other models. We also apply the pre-training task of CDR to other models to demonstrate its effectiveness.

\subsection{Experimental Setup}
\subsubsection{Datasets}

We experiment with public datasets \textbf{Mafengwo}, \textbf{Youshu}, and \textbf{Last-FM}. Mafengwo evaluates group recommendation models with tourist ratings for travel spots, while Youshu is used for bundle recommendations in book lists. Although Last-FM is not directly applicable to the tripartite graph-based recommendation, it can test CDR's generalization to other tasks due to its contained social relations. Experiments related to this dataset are given in Appendix \ref{ap:social}. Dataset statistics are given in Table \ref{tab:dataset}, and the meta-path types are given in Appendix \ref{ap:meta}.

\begin{table}
\caption{The statistics of the datasets.}
\label{tab:dataset}
\resizebox{0.4\textwidth}{!}{
\begin{tabular}{c|cccccc}

\hline
Dataset                  & \textbf{T}uple & \textbf{M}ember & \textbf{O}bject & \textbf{T}-\textbf{O} & \textbf{M}-\textbf{O} & \begin{tabular}[c]{@{}c@{}}Avg.\textbf{M}\\  per.\textbf{T}\end{tabular} \\ \hline
Mafengwo                 & 995   & 5,275  & 1,513  & 3,595                                                        & 39,761                                                        & 7.19                                                             \\ \hline
Youshu                   & 4,771 & 32,770 & 8,039  & 51,377                                                       & 138,515                                                       & 37.03                                                            \\ \hline
\multirow{2}{*}{Last-FM} & \textbf{U}ser  & \textbf{I}tem   & \multicolumn{2}{c}{\textbf{U}-\textbf{I}}                                       & \multicolumn{2}{c}{Social relation}                                                                                              \\ \cline{2-7} 
                         & 1,892 & 17,632 & \multicolumn{2}{c}{92,834}                                            & \multicolumn{2}{c}{25,434}                                                                                                       \\ \hline
\end{tabular}
}
\end{table}
\vspace{-1em}

\subsubsection{Baselines}

To validate the proposed CDR model, we juxtapose it with several top-tier baseline models across diverse recommendation domains, which are given below: 

\begin{itemize}[leftmargin=0.5cm]
\item \textbf{BPR} \cite{rendle2012bpr}: A classic user-oriented item ranking algorithm.
\item \textbf{LINE} \cite{tang2015line}: Embed nodes into low-dimensional space for recommendation.
\item \textbf{NGCF} \cite{wang2019neural}: Aggregate information in the interaction graph for collaborative filtering.
\item \textbf{SimGCL} \cite{yu2022graph}: Create contrastive views by adding noises to user/item embedding.
\item \textbf{XSimGCL} \cite{yu2023xsimgcl}: A method which generate contrastive views by employing a simple noise-based embedding augmentation.
\item \textbf{ALDI} \cite{huang2023aligning}: A cold-start method which transnfer warm items' behavioral information to cold items.
\item \textbf{BUIR} \cite{lee2021bootstrapping}: Uses only positive samples and data augmentation to alleviate data sparsity.
\item \textbf{DAM} \cite{chen2019matching}: Utilize attention to aggregate item embeddings for bundle recommendations.
\item \textbf{BGCN} \cite{9546546}: Unite user, item, and bundle nodes into a heterogeneous graph for information propagation.
\item \textbf{AGFN} \cite{li2023auto}: A bundle recommendation method which refines the neighborhood aggregation mechanism from different aspects.
\item \textbf{AGREE} \cite{cao2018attentive}: Determine group embedding fusion ratios using an attention mechanism.
\item \textbf{HCR} \cite{jia2021hypergraph}: Employ a two-channel hypergraph network for multi-task, group-user interest modeling.
\item \textbf{GroupIM} \cite{sankar2020groupim}: Maximize mutual information between groups and corresponding users.
\item \textbf{CrossCBR} \cite{10.1145/3534678.3539229}: Utilize and combines information from bundle and item views.
\item \textbf{ConsRec} \cite{wu2023consrec}: Explore consensus behind group behavior data.
\item \textbf{SBPR} \cite{zhao2014leveraging}: Integrate social relations to augment training samples for the model.
\item \textbf{DiffNet} \cite{wu2019neural}: Merge user representations from user-item bipartite graphs and social networks.
\item \textbf{SocialLGN} \cite{liao2022sociallgn}: Embed social relations into interaction graphs and executes graph convolution.
\end{itemize}

\subsubsection{Evaluation Metrics}

We utilize several metrics for evaluating recommendation performance, including F1-Score@$K$, Precision@$K$, Recall@$K$, and NDCG@$K$, which are prevalent in prior research \cite{he2020lightgcn, 9546546, jia2021hypergraph}.

\subsubsection{Implementation Details}

Addressing tripartite graph-based recommender systems' cold-start issue, this study enforces strict data constraints on models necessitating available tuple interactions. Models are trained using merely 5\% of tuple interactions, while 20\% are utilized for final performance assessment. Unlike previous work \cite{chen2019matching, cao2018attentive, jia2021hypergraph} that employs 100 random negative samples per test data, we apply Top-$K$ ranking on all dataset items/bundles, enhancing model evaluation accuracy. All models are set to a 64 embedding dimension to maintain fair comparison. Additionally, the early stopping method's patience is set at 10, halting training after 10 consecutive unimproved epochs. Model parameters are updated via Adam \cite{kingma2014adam}, with a 0.001 learning rate. Temperature hyperparameters $\tau$ for Mafengwo and Youshu datasets are configured to 1 and 0.3 for fine-tuning, and 3.8 and 1 for pre-training. For Last-FM dataset, $\tau$ is set to 1.2 (pre-training) and 3 (fine-tuning). 

\begin{table}[htbp]
\caption{Comparison of CDR and baselines on group and bundle recommendation datasets. \textbf{Bold} indicates best, and \underline{underlined} shows suboptimal results.}
\label{tab:performance comparison}
\centering
\resizebox{0.4\textwidth}{!}{
\begin{tabular}{c|ccccccc}
\hline
\textbf{Dataset}           & \textbf{Method} & \textbf{R@10}   & \textbf{R@20}   & \textbf{R@30}   & \textbf{N@10}   & \textbf{N@20}   & \textbf{N@30}   \\ \hline
\multirow{20}{*}{Mafengwo} & \multicolumn{7}{c}{Tuple Interactions}                                                                                      \\ \cline{2-8} 
                           & BPR             & 0.0604          & 0.1147          & 0.1682          & 0.0317          & 0.0460          & 0.0583          \\
                           & LINE            & 0.0243          & 0.0316          & 0.0566          & 0.0143          & 0.0164          & 0.0219          \\
                           & NGCF            & 0.0391          & 0.0795          & 0.1135          & 0.0159          & 0.0263          & 0.0334          \\
                           & LightGCN        & 0.0766          & 0.1384          & 0.1943          & 0.0385          & 0.0557          & 0.0682          \\
                           & BUIR            & 0.0969          & 0.1509          & 0.1928          & 0.0593          & 0.0740          & 0.0839          \\
                           & ColdNAS         & 0.0184          & 0.0234          & 0.0282          & 0.0214          & 0.0352          & 0.0501          \\
                           & SGL             & {\ul 0.1288}    & 0.2121          & 0.2626          & 0.0885          & {\ul 0.1088}    & {\ul 0.1212}    \\
                           & SimGCL          & 0.1263          & 0.2096          & 0.2500          & 0.0522          & 0.0740          & 0.0840          \\
                           & XSimGCL         & 0.0400          & 0.0643          & 0.0836          & {\ul 0.0988}    & 0.0929          & 0.0928          \\
                           & ALDI            & 0.1150          & {\ul 0.2460}    & {\ul 0.3203}    & 0.0554          & 0.0880          & 0.0999          \\
                           & CDR-F           & \textbf{0.1954} & \textbf{0.2806} & \textbf{0.3238} & \textbf{0.1268} & \textbf{0.1493} & \textbf{0.1594} \\
                           & Impro.          & 51.71\%         & 14.07\%         & 1.09\%          & 28.34\%         & 37.22\%         & 31.52\%         \\ \cline{2-8} 
                           & \multicolumn{7}{c}{Tuple Interaction/Member Interaction/Affiliation}                                                        \\ \cline{2-8} 
                           & AGREE           & 0.0388          & 0.0685          & 0.0994          & 0.0186          & 0.0268          & 0.0338          \\
                           & HCR             & 0.0559          & 0.0899          & 0.1219          & 0.0275          & 0.0327          & 0.0388          \\
                           & GroupIM         & 0.1453          & 0.2606          & 0.3170          & 0.0796          & 0.1109          & 0.1238          \\
                           & ConsRec         & {\ul 0.2312}    & {\ul 0.3085}    & {\ul 0.3487}    & {\ul 0.1256}    & {\ul 0.1449}    & {\ul 0.1534}    \\
                           & CDR             & \textbf{0.2694} & \textbf{0.3833} & \textbf{0.4309} & \textbf{0.1402} & \textbf{0.1715} & \textbf{0.1825} \\
                           & Impro.          & 16.52\%         & 24.25\%         & 23.57\%         & 11.62\%         & 18.36\%         & 18.97\%         \\ \hline
\multirow{19}{*}{Youshu}   & \multicolumn{7}{c}{Tuple Interactions}                                                                                      \\ \cline{2-8} 
                           & BPR             & 0.0188          & 0.0337          & 0.0442          & 0.0122          & 0.0170          & 0.0202          \\
                           & LINE            & 0.0234          & 0.0358          & 0.0467          & 0.0170          & 0.0208          & 0.0239          \\
                           & NGCF            & 0.0252          & 0.0479          & 0.0676          & 0.0167          & 0.0239          & 0.0298          \\
                           & LightGCN        & 0.0362          & 0.0573          & 0.0727          & 0.0269          & 0.0335          & 0.0379          \\
                           & BUIR            & {\ul 0.0508} & {\ul 0.0785}    & {\ul 0.1034}    & {\ul 0.0363}     & {\ul 0.0444}    & 0.0510          \\
                           & SGL             & 0.0247          & 0.0411          & 0.0530          & 0.0202          & 0.0257          & 0.0299          \\
                           & SimGCL          & 0.0313          & 0.0506          & 0.0602          & 0.0263          & 0.0322          & \textbf{0.0602} \\
                           & XSimGCL         & 0.0349          & 0.0581          & \textit{0.0734} & 0.0331          & 0.0401          & 0.0452          \\
                           & ALDI            & 0.0228          & 0.0290          & 0.0612          & 0.0159          & 0.0179          & 0.0199          \\
                           & CDR-F           & \textbf{0.0648} & \textbf{0.1000} & \textbf{0.1217} & \textbf{0.0398} & \textbf{0.0501} & {\ul 0.0563}    \\
                           & Impro.          & 27.41\%         & 27.32\%         & 17.78\%         & 9.59\%          & 12.74\%         & -               \\ \cline{2-8} 
                           & \multicolumn{7}{c}{Tuple Interaction/Member Interaction/Affiliation}                                                        \\ \cline{2-8} 
                           & DAM             & 0.0195          & 0.0316          & 0.0452          & 0.0134          & 0.0169          & 0.0214          \\
                           & CrossCBR        & 0.0281          & 0.0462          & 0.0611          & 0.0201          & 0.0262          & 0.0306          \\
                           & BGCN            & 0.0543          & 0.0848          & 0.1090          & 0.0358          & 0.0455          & 0.0528          \\
                           & AGFN            & {\ul0.0626} & {\ul 0.0981}    & {\ul 0.1247}    & {\ul 0.0457}    & {\ul 0.0565}    & {\ul 0.0652}    \\
                           & CDR             & \textbf{0.0903} & \textbf{0.1299} & \textbf{0.1520} & \textbf{0.0627} & \textbf{0.0744} & \textbf{0.0809} \\
                           & Impro.          & 44.25\%         & 32.42\%         & 21.89\%         & 37.20\%         & 31.68\%         & 24.08\%         \\ \hline
\end{tabular}}
\end{table}
\vspace{-1em}

\subsection{Performance Comparison}

To address \textbf{RQ1}, we assess models with highly sparse tuple interactions through experiments. Table \ref{tab:performance comparison} showcases the performance of CDR and baseline models, yielding the following analyses.

Firstly, CDR significantly surpasses baselines in group and bundle recommendations, averaging 18.88\% and 31.92\% improvement on the Mafengwo and Youshu datasets respectively. CDR-F, using only tuple interactions, remains highly competitive in general recommendation tasks. Notably, ConsRec and AGFN perform sub-optimally in extremely sparse settings, sometimes underperforming models like LightGCN and BUIR that solely utilize tuple interactions. The issue may stem from the inapplicability of multi-task joint optimization methods during the cold-start phase due to a severe data imbalance problem: with 5\% of tuple interactions for training amidst abundant member interactions. Thus, models like AGREE, HCR, and DAM, balancing tuple and member recommendations, struggle. ConsRec, by exploring consensus behind group behavior data, evades this imbalance issue, ranking as the best baseline in group recommendation. AGFN refines the aggregation mechanism to reach the best result in bundle recommendation. They still trail behind CDR.

Summarily, CDR’s pre-eminence arises chiefly due to: (1) The utility of consistency and discrepancy metrics, refining model learning of object-recommendee relationships. (2) The application of a pre-training\&fine-tuning approach, enhancing the generalization of our self-supervised task and adeptly sidestepping tuple-member interaction data imbalances. (3) The innovative contrastive CD loss, markedly unlocking the potential of quantified node relationships. We also conducted experiments about the performance of our CDR on social recommendation for detailed discussion and clarification on Last-FM dataset, which are given in Appendix \ref{ap:social}. Time complexity analysis is given in Appendix \ref{ap:time_analy}. 

\begin{table}[ht]

\caption{Performance comparison of CDR-P in extreme cold-start scenarios and baselines in warm-start scenarios.}
\label{tab:cold-start}
\centering
\resizebox{0.4\textwidth}{!}{
\begin{tabular}{c|cccc}
\hline
\multirow{2}{*}{Method} & \multicolumn{4}{c}{Mafengwo (Group)}                                  \\ \cline{2-5} 
                        & Recall@10       & Recall@20       & NDCG@10         & NDCG@20         \\ \hline
BPR                     & 0.0604          & 0.1147          & 0.0317          & 0.0460          \\
LINE                    & 0.0243          & 0.0316          & 0.0140          & 0.0164          \\
NGCF                    & 0.0391          & 0.0795          & 0.0159          & 0.0263          \\
LightGCN                & 0.0766          & 0.1384          & 0.0385          & 0.0557          \\
BUIR                    & {\ul 0.0969}    & {\ul 0.1509}    & {\ul 0.0593}    & {\ul 0.0740}    \\ \hline
CDR-P                   & \textbf{0.1275} & \textbf{0.1633} & \textbf{0.0653} & \textbf{0.0745} \\ \hline
                        & \multicolumn{4}{c}{Youshu (Bundle)}                                   \\ \cline{2-5} 
                        & Recall@10       & Recall@20       & NDCG@10         & NDCG@20         \\ \hline
BPR                     & 0.0188          & 0.0337          & 0.0122          & 0.0170          \\
LINE                    & 0.0234          & 0.0350          & 0.0170          & 0.0208          \\
NGCF                    & 0.0252          & 0.0479          & 0.0167          & 0.0239          \\
LightGCN                & {\ul 0.0362}    & {\ul 0.0573}    & {\ul 0.0269}    & {\ul 0.0335}    \\
BUIR                    & \textbf{0.0508} & \textbf{0.0785} & \textbf{0.0363} & \textbf{0.0444} \\ \hline
CDR-P                   & 0.0280          & 0.0476          & 0.0190          & 0.0251          \\ \hline
\end{tabular}
}
\vspace{-1em}
\end{table}

\begin{table}[htbp]
\vspace{-1em}
\centering
\caption{Performance of variants of CDR.}
\label{tab:Ablation}
\resizebox{0.4\textwidth}{!}{
\begin{tabular}{c|cccc}
\hline
\multirow{2}{*}{Method}      & \multicolumn{4}{c}{Mafengwo (Group)}                                  \\ \cline{2-5} 
                             & F1-Score        & Precision@20    & Recall@20       & NDCG@20         \\ \hline
CDR-P                        & 0.0195          & 0.0104          & 0.1633          & 0.0745          \\
CDR-F                        & 0.0336          & 0.0179          & 0.2806          & 0.1493          \\
CDR-R                        & 0.0425          & 0.0226          & 0.3531          & \textbf{0.1747} \\ \hline
w/o $c$                      & 0.0296          & 0.0157          & 0.2411          & 0.1160          \\
w/o $d$                      & 0.0349          & 0.0186          & 0.2790          & 0.1193          \\
w/o $c$\&$d$ & 0.0298          & 0.0158          & 0.2466          & 0.1177          \\ \hline
CDR                          & \textbf{0.0474} & \textbf{0.0253} & \textbf{0.3833} & 0.1715          \\ \hline
\multicolumn{1}{l|}{}        & \multicolumn{4}{c}{Youshu (Bundle)}                                   \\ \cline{2-5} 
\multicolumn{1}{l|}{}        & F1-Score        & Precision@20    & Recall@20       & NDCG@20         \\ \hline
CDR-P                        & 0.0135          & 0.0079          & 0.0476          & 0.0251          \\
CDR-F                        & 0.0207          & 0.0116          & 0.1000          & 0.0501          \\
CDR-R                        & 0.0171          & 0.0099          & 0.0621          & 0.0348          \\ \hline
w/o $c$                      & 0.0277          & 0.0157          & 0.1166          & 0.0648          \\
w/o $d$                      & 0.0179          & 0.0114          & 0.0423          & 0.0280          \\
w/o $c$\&$d$ & 0.0180          & 0.0114          & 0.0427          & 0.0281          \\ \hline
CDR                          & \textbf{0.0334} & \textbf{0.0191} & \textbf{0.1299} & \textbf{0.0744} \\ \hline
\end{tabular}}
\vspace{-1em}
\end{table}

\begin{figure*}[htbp]
	\centering
	\includegraphics[width=0.8\textwidth]{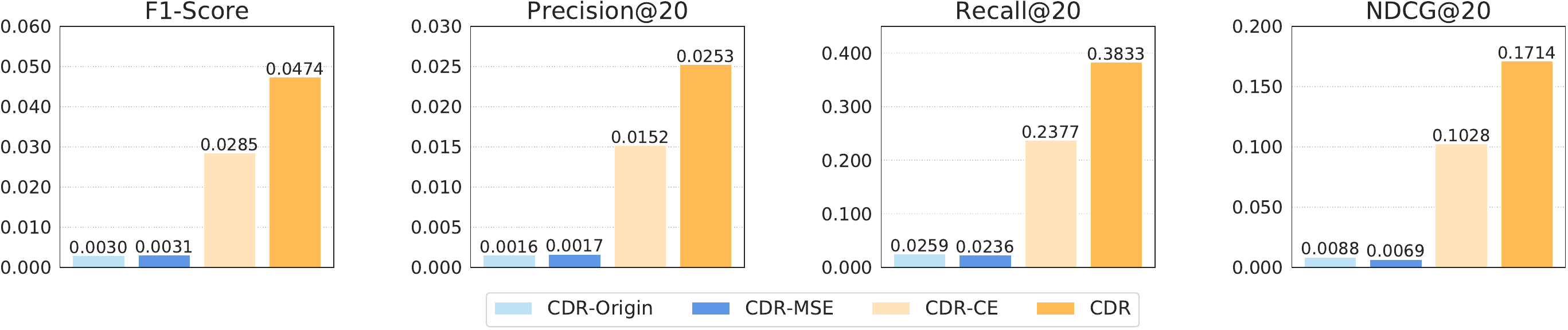}
	\caption{Performance comparison of different loss functions of CDR on the Mafengwo dataset.}
	\label{fig:loss}
\end{figure*}

\subsection{Performance in Extreme Cold-Start}

Addressing \textbf{RQ2}, we test CDR-P in extreme cold-start situations, with no tuple interactions available during training, contrasting with baselines using available tuple interactions. It is an unfair comparison since other models have more information from tuple interactions. Table \ref{tab:cold-start} reveals that our model, even without utilizing tuple interactions, outperforms all baselines on the Mafengwo dataset, showcasing CDR's robustness in extreme scenarios. The average performance of CDR-P in Youshu is the third for the lack of information.

CDR-P excels in extreme cold-start scenarios because its supervision signals, consistency and discrepancy metrics, derive from accessible and non-accessible meta-path views, built upon existing member interactions and affiliations. These member interactions and affiliations, being relatively abundant, furnish ample information for recommendations. While data noise could potentially derail model training, elevating the temperature hyperparameter concentrates on easily differentiated nodes, thereby boosting robustness against noisy datasets. Consequently, CDR-P discerns approximate node preferences, mitigating the adverse effects of data noise.

\subsection{Ablation Study} 
Addressing \textbf{RQ3}, ablation tests on CDR components confirm each part's significance. CDR-P and CDR-F denote models using only pre-training and solely tuple interactions, respectively, while CDR-R swaps pre-training and fine-tuning data—using tuple interactions for pre-training and member ones for fine-tuning. Table \ref{tab:Ablation} reveals omitting pre-training or fine-tuning stages reduces performance, affirming CDR's pre-training and fine-tuning efficacy. Swapping training data stages also hampers performance, with a notable degradation on the Youshu dataset after fine-tuning via member interactions and affiliations, indicating the pivotal role and ordering of tuple interactions in tripartite graph-based recommendations.

\begin{figure*}[h]
	\centering
	\includegraphics[width=0.8\textwidth]{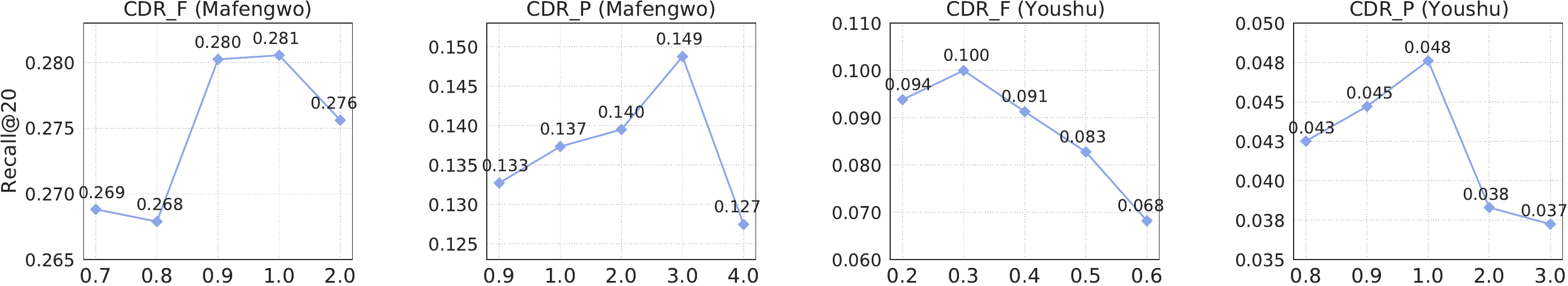}
	\caption{The effect of different temperatures $\tau$ on the pre-training and fine-tuning phases.}
	\label{fig:t}
\end{figure*}

\begin{figure*}[]
	\centering
	\includegraphics[width=0.8\textwidth]{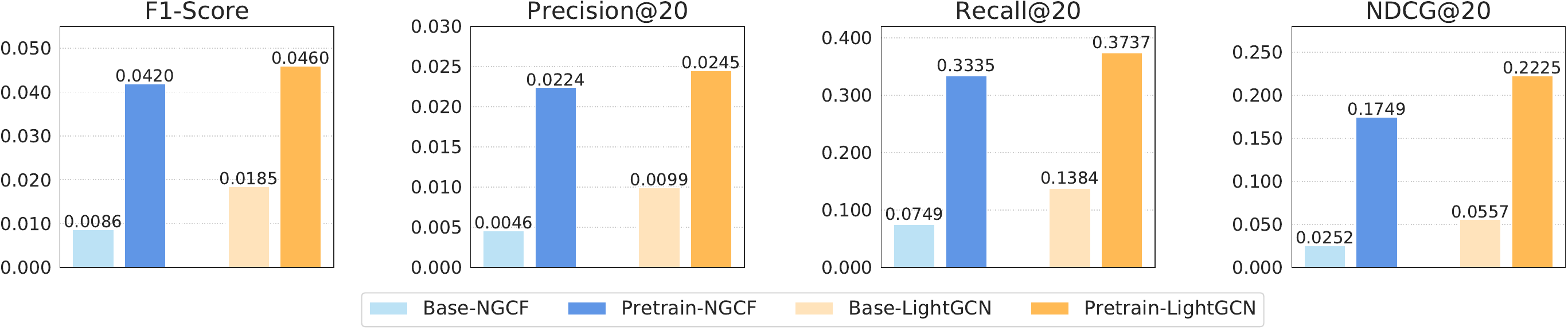}
	\caption{The performance comparison of the two baselines on the Mafengwo dataset before and after CDR-P pre-training.}
	\label{fig:pre-train}
\end{figure*}

\subsubsection{Ablation on Consistency and Discrepancy Metrics}
Our consistency and discrepancy metrics, reflecting complex tuple-object node relationships beyond binary labels, embody real-life varied user preferences for numerous purchased items. Validating these metrics’ effectiveness, we binarize calculated consistency $c$ and discrepancy $d$. Table \ref{tab:Ablation} reveals substantial performance drops when omitting either metric, affirming their complementarity in contrastive learning-based loss and thus their collective necessity.

\subsubsection{Ablation on the CD Loss}

The CDR model's loss function was substituted with three alternatives to validate the proposed contrastive learning-inspired CD loss. Specifically, CDR-Origin employs the original loss (Eq. (\ref{equation:combine_loss})), while CDR-MSE and CDR-CE utilize mean squared error \cite{hu2008collaborative} and cross-entropy loss \cite{he2017neural} respectively. Results in Fig. \ref{fig:loss} reveal CDR-Origin and CDR-MSE underperform compared to CDR, potentially due to information loss from the subtraction between consistency and discrepancy, as discussed in Section \ref{sec:motivation}. 

The performance of CDR-CE and CDR, which use these node relationships separately, is better than CDR-Origin and CDR-MSE. The CD loss of CDR performed best due to its characteristics to distinguish relatively easy or difficult nodes by the temperature $\tau$. CD loss can adapt to data containing noise of different degrees.

\subsection{Hyperparameter Sensitivity Analysis}
To further answer \textbf{RQ4}, we conducted hyperparameter sensitivity experiments on CDR, where Fig. \ref{fig:t} shows the effect of different temperature values $\tau$ in the pre-training and fine-tuning phases of Mafengwo and Youshu datasets. We can find that CDR is sensitive to the temperature $\tau$, and the optimal temperature value in the pre-training stage is higher than that in the fine-tuning stage on these two datasets. This occurs because member interactions are noisier than tuple interactions. 

Therefore, a lower temperature is conducive to distinguishing similar nodes, enabling the learning of fine-grained node representations for tuple recommendations. Furthermore, an analysis concerns the correlation between consistency \& discrepancy and loss is given in Appendix \ref{ap:correlation}.

\subsection{Generality of the Pre-Training Task}
Finally, we use the consistency and discrepancy derived in the pre-training phase of CDR as the supervision signal to train two traditional recommender baselines to verify the generalization ability of the two metrics. In Fig. \ref{fig:pre-train}, \textit{Base} and \textit{Pretrain} represent the performance of the baselines with/without metrics obtained from CDR-P pre-training as supervision signals, respectively. From Fig. \ref{fig:pre-train}, we can find that, with consistency and discrepancy as the supervision signal, the performance of NGCF and LightGCN improves substantially compared to the models that use only binary tuple interactions. The results further demonstrate the importance of member interactions and member affiliations and the effectiveness of the derived two metrics.

\section{Conclusions}
\label{section:conclusion}
We introduce CDR, a novel tripartite graph-based recommendation model to improve its performance in cold-start scenarios. The model utilizes two new metrics, consistency and discrepancy, representing tuple-object relationships via reachable/non-reachable meta-paths, based on ample user-item interactions and group-user/bundle-item affiliations. Leveraging the GCN's limit theory, these metrics are efficiently pre-calculated before optimization. Furthermore, a novel contrastive learning-based objective is proposed, independently exploiting consistency and discrepancy as positive and contrastive supervision signals. Extensive experiments across various recommendation tasks and multiple real-world datasets confirm the proposed metrics and contrastive learning-inspired CD loss's generalization and robustness to tripartite graph-based recommendation's cold-start issues.

\section*{Acknowledgement}
We thank the anonymous reviewers for their insightful feedback and constructive comments. This work was partially supported by the National Natural Science Foundation of China (Grant No. 6217602), the Australian Research Council through the Industrial Transformation Training Centre (Grant No. IC200100022), and the Science and Technology Research Program of Chongqing Municipal Education Commission (Grant No. KJZD-K202204402 and KJZD-K202304401). Co-authors Yaochen Zhu and Chen Chen consulted on this project on unpaid weekends for personal interests, and appreciated collaborators and family for their understanding.

\bibliographystyle{ACM-Reference-Format}
\balance
\bibliography{sample-base}

\appendix

\section{Related Work}
\label{ap:rela}
\subsection{Tripartite Graph-Based Recommendation}
In conventional recommendation systems, user-item interactions are typically represented as a bipartite graph, with various Graph Convolutional Network (GCN)-based methods modeling these interactions. This paper introduces a novel concept: tripartite graph-based recommendations, which entail recommendation tasks involving interactions among three distinct entity types. The third entity often comprises user and item groups, introducing a novel affiliation between items and bundles or users and groups. In this model, if an item (recommended object) is recommended to a user group (recommendee), it is termed a “group recommendation task” \cite {9195784, 8613823, 8663390}; if an item group (recommended object) is recommended to a user (recommendee), it is known as a “bundle recommendation” \cite{9546546,chen2019matching, 10.1145/3534678.3539229}. As the first to consolidate these into one overarching problem, we delineate group and bundle recommendations separately in the ensuing section.

\subsubsection{Group Recommendation}
Group recommendation targets recommending a single item to a user group, accounting for varied interests. Often formed temporarily, user groups usually yield sparse group-item interactions\cite{9195784,sankar2020groupim,zhang2021double}. Guo et al. \cite{guo2020group} introduced a model addressing sparsity by leveraging a self-attention mechanism to learn users' social influence in temporary group recommendations. Similarly, Cao et al. \cite{cao2018attentive} proposed AGREE, merging neural collaborative filtering and attention mechanisms to dynamically determine group embedding fusion weights. Jia et al. \cite{jia2021hypergraph} advanced this by fusing member and group networks into a dual-channel hypergraph convolutional model, HCR.

\subsubsection{Bundle Recommendation}
Bundle recommendation seeks to suggest multiple items collectively to a single user and, akin to group recommendation, faces a pervasive data sparsity issue due to the arbitrary bundling of items. Attention networks are often utilized to ascertain the weight of items in a bundle \cite{10.1145/3534678.3539229}. Furthermore, multi-task learning is commonly employed to meld item and bundle interaction data, mitigating sparsity. For instance, Chen et al. \cite{chen2019matching} devised DAM, a bundle recommendation model that interlinks user-bundle and user-item interactions within a multi-task joint optimization framework. Gao et al. \cite{9546546} amalgamated user-item interactions, user-bundle interactions, and bundle-item affiliations into a heterogeneous graph to ensure bundle embeddings encapsulate item semantics.

\subsection{Pre-Training and Fine-Tuning in Recommendation}

\textbf{Pre-training}, aiming to optimize model parameters utilizing label ground truth or auxiliary data, facilitates faster convergence or reduces label needs during subsequent \textbf{fine-tuning} stages \cite{liu2021pre,wang2021pre,10.1145/3568953,hao2023multi}. In recommendations, pre-training and fine-tuning, viewed as self-supervised learning, entails initial training on augmented data, followed by fine-tuning on original data, adapting to the designated recommendation task.
 
Key aspects of pre-training involve creating augmented data and devising a pretext task. Chen et al. \cite{chen2019bert4sessrec} and Zhang et al. \cite{zhang2021unbert}, inspired by the NLP model BERT \cite{devlin2018bert}, designed similar models for recommendations where side information on users/items is available. Wang et al. \cite{wang2023curriculum} leveraged this by viewing interaction data through heterogeneous graphs and building specific subgraphs. Similarly, models utilizing side information in pre-training have been proposed \cite{zhou2020s3}. Qiu et al. \cite{qiu2021u} masked sections of user and item reviews, using bidirectional context to reconstruct them.

\section{Algorithm}
\label{ap:algo}
The overall process of CDR is presented in Algorithm \ref{alg:algorithm}. 
\begin{algorithm}[h]
	\caption{The overall process of CDR.}
	\LinesNumbered 
	\label{alg:algorithm}
	\KwIn{Tuple interaction matrix $Y$ \qquad \qquad \qquad \qquad \qquad \qquad \qquad \qquad \qquad \qquad \qquad Member interaction matrix $X$ \qquad \qquad \qquad \qquad \qquad \qquad \qquad \qquad \qquad \qquad Tuple-member affiliation matrix $Z$ }
	\KwOut{Final node embeddings $\mathbf{E}$}
	\DontPrintSemicolon
	Extract $\boldsymbol{C}_{member}$ and $\boldsymbol{D}_{member}$ from member interaction $X$ and affiliation $Z$\\
	Randomly initialize embedding $\mathbf{E}^{p}$\\
	\While{not converge}{
	    \For{each epoch}{
			Evaluate $\mathcal{L}$ through $\boldsymbol{C}_{member}$, $\boldsymbol{D}_{member}$, and $\mathbf{E}^{p}$ (Eq. (\ref{equation:contrastive}))\\
    		Backpropagation and update embeddings $\mathbf{E}^{p}$\\
	    }
	}
	Extract $\boldsymbol{C}_{tuple}$ and $\boldsymbol{D}_{tuple}$ from tuple interactions $Y$\\
	Stop gradient of $\mathbf{E}^{p}$\\
	Initialize node embedding $\mathbf{E}^{f}$ with $\mathbf{E}^{p}$\\
	Concatenate final embeddings $\mathbf{E} \gets \mathbf{E}^{f} \parallel \mathbf{E}^{p}$ \\
	\While{not converge}{
	    \For{each epoch}{
			Evaluate $\mathcal{L}$ through $\boldsymbol{C}_{tuple}$, $\boldsymbol{D}_{tuple}$, and $\mathbf{E}$ (Eq. (\ref{equation:contrastive}))\\
    		Back propagation and update node embeddings $\mathbf{E}^{f}$\\
	    }
	}
    \textbf{return} Final node embeddings $\mathbf{E}$\\
\end{algorithm}

\section{Supplementary Experiments}

\label{ap:social}

To further explore if CDR can generalize to other recommendation types with existing consistency and discrepancy metrics. we validate CDR's adaptability to other recommendation tasks, testing on the Last-FM social recommendation dataset. The baseline methods we choose are \textbf{SBPR} \cite{zhao2014leveraging}, \textbf{DiffNet} \cite{wu2019neural}, and  \textbf{SocialLGN} \cite{liao2022sociallgn}. As Table \ref{tab:last-fm} indicates, models utilizing social relations outperform those relying solely on interaction data. CDR-F remains competitive using only user-item interactions. In social recommendation, which only involves user social relationships, CDR-P outperforms CDR-F by leveraging both social and interaction data. Post-fine-tuning, CDR, amalgamating the benefits of CDR-P and CDR-F, enhances all metrics, substantiating its generalizability across various recommendation domains where a tripartite graph can be constructed.

\begin{table}[]

\caption{Performance comparison of CDR model and baselines under social recommendation scenarios.}
\label{tab:last-fm}
\centering
\resizebox{0.45\textwidth}{!}{
\begin{tabular}{c|ccccc}
\hline
Dataset                   & Method    & F1-Score        & Precision@20       & Recall@20          & NDCG@20            \\ \hline
\multirow{12}{*}{Last-FM} & \multicolumn{5}{c}{User-item interaction}                                         \\ \cline{2-6} 
                          & BPR       & 0.0277          & 0.0207          & 0.0417          & 0.0353          \\
                          & BUIR      & 0.0484          & 0.0361          & 0.0731          & 0.0685          \\
                          & NGCF      & 0.0505          & 0.0379          & 0.0756          & 0.0721          \\
                          & LightGCN  & {\ul 0.0568}    & {\ul 0.0425}    & {\ul 0.0857}    & \textbf{0.0844} \\ \cline{2-6} 
                          & CDR-F     & \textbf{0.0576} & \textbf{0.0428} & \textbf{0.0878} & {\ul 0.0798}    \\ \cline{2-6} 
                          & \multicolumn{5}{c}{User-item interaction/Social relation}                         \\ \cline{2-6} 
                          & SBPR      & 0.0544          & 0.0405          & 0.0829          & 0.0725          \\
                          & DiffNet   & 0.0570          & 0.0427          & 0.0859          & 0.0805          \\
                          & SocialLGN & 0.0665          & 0.0495          & 0.1011          & 0.0894          \\ \cline{2-6} 
                          & CDR-P     & {\ul 0.0756}    & {\ul 0.0563}    & {\ul 0.1151}    & {\ul 0.1108}    \\
                          & CDR       & \textbf{0.0836} & \textbf{0.0622} & \textbf{0.1274} & \textbf{0.1195} \\ \hline
\end{tabular}}
\vspace{-1em}
\end{table}

\section{Time Complexity Analysis}
\label{ap:time_analy}
Our analysis covers each critical stage of our method, i.e., pre-processing, pre-training, and fine-tuning.
In the pre-processing stage, for each member $m$, we need to calculate two kinds of terms, i.e., $\delta_{m,t}$ and $\delta_{m,o}$, and the time complexity is $O(|\mathcal{M}||\mathcal{T}|)+O(|\mathcal{M}||\mathcal{O}|)$, where $\mathcal{T,O,M}$ are sets of the three types of nodes in a tripartite graph-based recommendation, and $||$ is the number of elements in a set. Similarly, we need to calculate two other kinds of terms $\delta_{o,t}$ and $\delta_{t,o}$, and the time complexity is $O(\mathcal{|O||T|})$. Then, with the four kinds of terms, i.e., $\delta_{m,t}$, $\delta_{m,o}$, $\delta_{o,t}$, and $\delta_{t,o}$, we can directly get the values in $C_{member}$, $D_{member}$, $C_{tuple}$, and $D_{tuple}$ for further training. All calculations in pre-processing do not include matrix multiplication, so this stage does not take a lot of time.

In our model, the time complexity for both the pre-training and fine-tuning stages is similar. Here, $Q$ represents the set of node pairs that exhibit positive consistency and whose types belong to tuples or objects. According to the loss function, we can derive the time complexities for every batch of the two stages are both $O((\frac{|Q|}{B}+|Q|)d)$, where $d$ is the dimension of embedding, and $B$ is the batch size. Note that a pair of nodes is included in the set $Q$ only if they share common neighborhood nodes. Given the sparsity characteristic of recommendation datasets, $Q$ is significantly smaller than $\mathcal{O\times T}$. This inherent sparsity greatly reduces the actual computational time, thereby ensuring efficiency in processing.

We conduct a comparison of model runtime, with the results shown in Table \ref{tab:runtime}. Our CDR is the fastest because it does not require node aggregation. ALDI \cite{huang2023aligning} and XSimGCL \cite{yu2023xsimgcl}, which only utilize T-O interactions, also offer relatively fast speeds. In contrast, AGFN \cite{li2023auto} processes more slowly for using graph aggregation methods on tripartite graph.

\begin{table}[]
\caption{The runtime of different graph-based method.}
\label{tab:runtime}
\resizebox{0.35\textwidth}{!}{
\centering
\begin{tabular}{|l|l|l|l|l|}
\hline
Method  & ALDI    & XSimGCL & AGFN    & CDR     \\ \hline
Runtime & 0.0820s & 0.0901s & 0.2263s & 0.0738s \\ \hline
\end{tabular}}
\vspace{-1em}
\end{table}

\section{Types of Meta-paths}
\label{ap:meta}
We summarize all the meta-path types in Mafengwo and Youshu datasets, and the result is shown in Table \ref{tab:metapath}, where G indicates a group, B indicates a bundle, and U and I are a user and an item separately.

\begin{table}[h]
\caption{The metapaths for different scenarios.}
\label{tab:metapath}
\centering
\begin{tabular}{|c|c|cc|}
\hline
Scenarios                & Dataset                  & \multicolumn{2}{c|}{Meta-path}                   \\ \hline
\multirow{2}{*}{Group} & \multirow{2}{*}{Mafengwo} & \multicolumn{1}{l|}{Pre-train} & GUG/GUI/IUG/IUI \\
                         &                          & \multicolumn{1}{l|}{Fine-tune} & GIG/GI/IG/IGI   \\ \hline
\multirow{2}{*}{Bundle}  & \multirow{2}{*}{Youshu}  & \multicolumn{1}{l|}{Pre-train} & UIU/UIB/BIU/BIB \\
                         &                          & \multicolumn{1}{l|}{Fine-tune} & UBU/UB/BU/BUB   \\ \hline
\multirow{2}{*}{General} & Mafengwo                 & \multicolumn{2}{c|}{GIG/GI/IG/IGI}               \\
                         & Youshu                   & \multicolumn{2}{c|}{UBU/UB/BU/BUB}               \\\hline
\end{tabular}
\end{table}
\vspace{-1em}

\section{Correlation between Consistence \& Discrepancy and the Training Loss}
\label{ap:correlation}
We conduct an experiment by calculating the Pearson correlation coefficients between the metrics consistency ($c$) and discrepancy ($d$), and the loss for each node pair, as presented in Table \ref{tab:correlation}. The experimental results indicate that there is a negative correlation between consistency and loss, and a positive correlation between discrepancy and loss. This means node pairs with high consistency tend to have lower loss, while those with greater discrepancy exhibit higher loss.

\begin{table}[]
\vspace{-1em}
\caption{The Pearson correlation coefficients between the metrics consistency ($c$) and discrepancy ($d$), and the loss for each node pair.}
\label{tab:correlation}
\centering
\begin{tabular}{|c|c|c|}
\hline
Dataset  & Corr($c$,loss) & Corr($d$,loss) \\ \hline
Youshu   & -0.60        & 0.52         \\ \hline
Mafengwo & -0.46        & 0.43         \\ \hline
\end{tabular}
\vspace{-1em}
\end{table}

This pattern of correlations allows us to leverage the metrics $c$ and $d$ strategically during the optimization process. To effectively minimize loss, an increase in $c_{v_i,v_j}$ should correspond with an increase in $cos(e_{v_i}, e_{v_j})$, enhancing similarity between embeddings. Conversely, an increase in $d_{v_i,v_j}$ should result in a decrease in $cos(e_{v_i}, e_{v_j})$, reducing similarity. 
\end{document}